\documentclass[11pt]{article}
\usepackage{latexsym}  
\usepackage{amssymb}
\usepackage{graphicx}
\usepackage{amsmath}

\begin{document}


\hspace{8cm}{OU-HET-730/2011; MISC-2011-16}

\vspace{3mm}

\begin{center}
{\Large\bf SU(5) Compatible Yukawaon Model \\
With Two Family Symmetries U(3)$\times$O(3)}

\vspace{3mm}
{\bf Yoshio Koide}

{\it Department of Physics, Osaka University,  
Toyonaka, Osaka 560-0043, Japan} \\
{\it E-mail address: koide@het.phys.sci.osaka-u.ac.jp}

\date{\today}
\end{center}

\vspace{3mm}
\begin{abstract}
A yukawaon model which is compatible
with an SU(5) GUT model is investigated.
In a previous SU(5) compatible yukawaon model 
with a U(3) family gauge symmetry, 
we could not build a model with a lower energy scale of 
the family gauge symmetry breaking scale $\Lambda_{fam}$ 
than $10^{13}$ GeV, so the family gauge boson effects in the 
previous model were invisible.
In the present model, we consider two family symmetries 
U(3)$\times$O(3), and we assume that the conventional quarks 
and leptons $(\bar{\bf 5}+{\bf 10}+{\bf 1})$ of SU(5)  
are described as
$(\bar{\bf 5}_i+{\bf 10}_\alpha+{\bf 1}_\alpha)$ 
($ i=1,2,3$ and $\alpha=1,2,3$ are indices of U(3) and 
O(3), respectively).
As a result, we build a model with $\Lambda_{O3}
\sim 10^{16}$ GeV and $\Lambda_{U3} 
\sim 10^{3}$ GeV.
The lightest U(3) family gauge boson $A_1^1$ 
will be observed with a mass of the order of 1 TeV.
\end{abstract}

\vspace{5mm}

{\large\bf 1. Introduction}

In the standard model (SM) of quarks and leptons, their mass spectra 
and mixings originate in the structures of the Yukawa coupling
constants, although the masses themselves originate in the Higgs scalar.
The Yukawa coupling constants are fundamental constants in the 
theory, so that they are not quantities which we can evaluate dynamically. 
If we intend to understand the observed mass spectra and mixings by 
a ``family symmetry", we cannot adopt a non-Abelian gauge symmetry, 
because the Yukawa coupling constants 
play a role in breaking the symmetry.
Of cause, instead of such a non-Abelian symmetry, we may assume U(1) 
symmetries, discrete symmetries, and so on. 
Then, by requiring that the model is invariant under such a symmetry,
we can obtain some constraints on the Yukawa coupling constants.
However, even if we consider such symmetries, we still have a trouble 
\cite{no-go}: 
We know that any model with a family symmetry cannot derive 
a realistic flavor mixing matrix (Cabibbo-Kobayasi-Maskawa \cite{CKM}
(CKM) quark mixing matrix and/or Pontecorvo-Maki-Nakagawa-Sakata 
\cite{PMNS} (PMNS) lepton mixing matrix) unless we do not consider 
a multi-Higgs model.
However, such the multi-Higgs model usually lead to a flavor 
changing neutral current (FCNC) problem.

An easy way to escape from these problems is to 
consider that the mass spectra and mixings originate in
vacuum expectation values (VEVs) of new scalars.
As one of such models, the so-called ``yukawaon" model 
\cite{yukawaon} is known.
In the yukawaon model, which is a kind of 
``flavon" model \cite{flavon},  all effective 
Yukawa coupling constants $Y_f^{eff}$  ($f=u,d,e,\cdots$) 
are given by VEVs of ``yukawaons" $Y_f$ as 
$$
(Y_f^{eff})_{ij} = \frac{y_f}{\Lambda} \langle (Y_f)_{ij} \rangle ,
\eqno(1.1)
$$
that is, would-be Yukawa interactions are given by 
the following superpotential:
$$
W_Y = \frac{y_e}{\Lambda} {\ell}_i Y_e^{ij} e_j^c H_d 
+ \frac{y_\nu}{\Lambda} {\ell}_i Y_\nu^{ij}\nu_j^c H_u 
+\lambda_R \nu_i^c Y_R^{ij} \nu_j^c 
+ \frac{y_u}{\Lambda} u_i^{c} Y_u^{ij} q_j H_u 
+ \frac{y_d}{\Lambda} d_i^c Y_d^{ij} q_j H_d ,
\eqno(1.2)
$$
where $\ell$ and $q$ are SU(2)$_L$ doublets 
$\ell=(\nu_L, e_L)$ and $q=(u_L, d_L)$.
In order to distinguish each yukawaon from others,
$Y_f$ have $R$ charges different from each other, 
and we assume $R$ charge conservation.
(Of course, the $R$ charge conservation is broken
at a high energy scale $\Lambda_{fam}$ at which the
family symmetry is broken.)

The most notable characteristic of the yukawaon model is 
that structures of VEV matrices $\langle Y_f \rangle$  
are described in terms of only one fundamental VEV matrix  
$$
\langle \Phi_e \rangle = k_0 \, {\rm diag}(\sqrt{m_e},
\sqrt{m_\mu}, \sqrt{m_\tau}) .
\eqno(1.3)
$$
For examples, we describe $\langle Y_f \rangle$ as follows
in terms of $\langle \Phi_e \rangle$ \cite{O3_09PLB}:
$$
\langle Y_e\rangle = k_e \langle \Phi_e \rangle
\langle \Phi_e \rangle,
\eqno(1.4)
$$
$$
\langle Y_u \rangle = k_u \langle \Phi_u \rangle
\langle \Phi_u \rangle ,  \ \ \ \ 
\langle \Phi_u \rangle = k'_u \langle \Phi_e \rangle
({\bf 1} + a_u X) \langle \Phi_e \rangle ,
\eqno(1.5)
$$
$$
\langle Y_d \rangle = k'_d \langle \Phi_e \rangle
({\bf 1} + a_d X) \langle \Phi_e \rangle ,
\eqno(1.6)
$$
where
$$
{\bf 1} = \left(
\begin{array}{ccc}
1 & 0 & 0 \\
0 & 1 & 0 \\
0 & 0 & 1 
\end{array} \right) , \ \ \ 
X= \frac{1}{3} \left(
\begin{array}{ccc}
1 & 1 & 1 \\
1 & 1 & 1 \\
1 & 1 & 1 
\end{array} \right) . 
\eqno(1.7)
$$
We can also describe the neutrino mass matrix $M_\nu$ 
in terms of $\langle \Phi_e \rangle$ (see Eqs.(2.19)
and (3.20) later).   
In this scenario, we do not ask why the VEV matrix
$\langle \Phi_e \rangle$ takes such a value given in 
Eq.(1.3). 
As a result, the model has considerably few adjustable 
parameters (the charged lepton masses are input values, 
and the eigenvalues of $\langle \Phi_e \rangle$ are not
adjustable parameters). 
The observed hierarchical structures in quarks and leptons
are attributed to the hierarchical structure of 
$\langle \Phi_e \rangle$. 

Here, note that the yukawaons $Y_f$ are singlets under the conventional
gauge symmetries SU(3)$_c \times$ SU(2)$_L\times$U(1)$_Y$,
and they have only family indices.
This suggests that the yukawaon model may be compatible with a 
grand unification (GUT) model, for example, SU(5) GUT model \cite{SU5}.
Recently, the author \cite{SU5_1110} has proposed an SU(5) compatible 
yukawaon model.
The main purpose of the SU(5) compatible model was to build a yukawaon model
without a cutoff scale $\Lambda$.
The purpose also was to develop the yukawaon model and not to discuss 
problems in a GUT model. 
That is, possible structures of yukawaons were investigated 
for the case when we regarded quarks and leptons as 
$\bar{\bf 5}+{\bf 10}+{\bf 1}$ of SU(5).
In the present paper, too, a compatible SU(5) yukawaon model is 
investigated, 
but we do not intend to develop a GUT scenario or to resolve problems
in the current GUT scenarios.

Let us give a brief review of the previous SU(5) compatible 
yukawaon model \cite{SU5_1110} in order to make the purpose of the present 
paper clear.
In the previous model, superpotential terms for up-quark and charged 
lepton yukawaon sectors have been taken as:
$$
W_{Yu} = y_{u} {\bf 10}_i Y_u^{ij} \overline{\bf 10}_j^{\prime} 
+M_{10} \overline{\bf 10}_i^{\prime}  {\bf 10}^{\prime\, i}
+ y_{10} {\bf 10}^{\prime\, i} {\bf 10}_i {\bf 5}_H ,
\eqno(1.8)
$$
$$
W_{Ye} = y_{e} \bar{\bf 5}_i Y_e^{ij} {\bf 5}_j^{\prime} 
+M_{5} {\bf 5}_i^{\prime} \bar{\bf 5}^{\prime\, i}  
+ y_{5} \bar{\bf 5}^{\prime \, i} {\bf 10}_i 
\bar{\bf 5}_H ,
\eqno(1.9)
$$
which lead to effective Yukawa interactions
$$
W_{Yu}^{eff} = \frac{y_u y_{10}}{\bar{M}_{10}} {\bf 10}_i Y_{u}^{ij} 
{\bf 10}_j {\bf 5}_H ,
\eqno(1.10)
$$
$$
W_{Ye}^{eff} = \frac{y_e y_{5}}{\bar{M}_{5}} \bar{\bf 5}_i Y_{e}^{ij} 
{\bf 10}_j \bar{\bf 5}_H,
\eqno(1.11)
$$
respectively.
Here, although $\bar{M}_{10}$ and $\bar{M}_5$ in Eqs.(1.10) and (1.11) 
have family-number dependence as we discuss later, for the time being
those may be regarded as  $\bar{M}_{10} \simeq M_{10}$ and 
$\bar{M}_5 \simeq M_5$.
Anyhow, as seen in Eqs.(1.10) and (1.11), 
we can introduce two different cutoff scales $M_{10}$
and $M_5$ for the up-quark and charged lepton sectors, respectively.
However, since the model gives 
$$
M_u = \frac{y_u y_{10}}{M_{10}} \langle Y_u^{ij} \rangle v_{Hu},
\ \ \ 
M_e = \frac{y_e y_{5}}{M_{5}} \langle Y_{e}^{ij} \rangle v_{Hd} ,
\eqno(1.12)
$$
where $v_{Hu} = \langle H_u^0 \rangle=\langle {\bf 5}_H\rangle$
and $v_{Hd} = \langle H_d^0 \rangle=\langle \bar{\bf 5}_H \rangle$,
we are obliged to accept phenomenological constraints  
$$
\frac{ \langle Y_u \rangle }{M_{10}} \sim 1, 
\ \ \ 
\frac{\langle Y_{e} \rangle}{M_5} \sim 10^{-1} ,
\eqno(1.13)
$$
from the observed quark and lepton masses (we 
suppose $\tan\beta \sim 10$).
(Here, the order of a VEV matrix $\langle Y_f \rangle$ 
means the largest value of the eigenvalues of 
$\langle Y_f \rangle$.)
We consider that the family symmetry U(3) 
is broken at an energy scale $\Lambda_{U3}$.
The scale is given by the largest one of the VEV values of 
the whole U(3) non-singlet scalars, i.e.
$\Lambda_{U3} \ge \langle Y_u \rangle$.
If we want that the family symmetry effects are visible, 
we must take the value of $\Lambda_{U3}$ considerably low. 
However, on the other hand, if we take $M_{10} < 10^{12}$ GeV, 
such a model with a low value of $M_{10}$ will cause 
blowing up of the SU(3)$_c$ gauge coupling constant 
because of the additional fields 
$({\bf 10}'+\overline{\bf 10}')$. 
In order to avoid such the blowing up, we must take
$M_{10} \ge 10^{12}$ GeV.
Thus, the scale $\Lambda_{U3}$ is constrained as 
$$
\Lambda_{U3} \ge \langle Y_u \rangle \sim M_{10} \ge 
 10^{12} \ {\rm GeV} .
 \eqno(1.14)
$$
We could not take a lower value of $\Lambda_{U3}$ in the 
previous SU(5) compatible model \cite{SU5_1110}.

The main purpose of the present paper is to propose an SU(5)-compatible
yukawaon model in which the family symmetry U(3) is broken at a suitably 
low energy scale $\Lambda_{U3} \sim 10^3$ GeV.
The basic idea is quite simple:
in the conventional quarks and leptons $\bar{\bf 5}+{\bf 10}$
of SU(5), 
the field $\bar{\bf 5}$ is ${\bf 3}$ of the family symmetry U(3)
[we denote it as $\bar{\bf 5}_i$ ($i=1,2,3)$], while the field 
${\bf 10}$ is ${\bf 3}$ of another family symmetry O(3) [we denote 
it as ${\bf 10}_\alpha$ ($\alpha = 1,2,3$)].
Thereby, VEVs of the yukawaons $Y_e$ and $Y_u$ are given by
$\langle Y_e^{i\alpha}\rangle$ and $\langle Y_u^{\alpha\beta}\rangle$,
so that we can assume that those VEV values take different scales
$\langle Y_e^{i\alpha}\rangle \sim \Lambda_{U3}$ and 
$\langle Y_u^{\alpha\beta}\rangle \sim \Lambda_{O3}$, 
where $\Lambda_{U3}$ and $\Lambda_{O3}$ are energy scale 
at which U(3) and O(3) are broken, respectively.
We consider $\Lambda_{O3} \gg \Lambda_{U3}$.
[A model with two family symmetries U(3)$\times$O(3) has 
been proposed by Sumino \cite{Sumino09}.
A yukawaon model with two family symmetries U(3)$\times$O(3) 
has been discussed in Ref.\cite{U3O3_11JPG}, although 
the model was not compatible with SU(5).]

In addition to the above idea, we will propose the following new
ideas in the present yukawaon model:

(i) {\it Economizing of yukawaons}: 
In the previous SU(5) compatible yukawaon model \cite{SU5_1110}, 
we have demonstrated 
that the yukawaon $Y_\nu$ in Eq.(1.2) can be substituted with 
the charged lepton yukawaon $Y_e$.
In the present model, the up-quark yukawaon $Y_u$ will also be 
removed from the model by modifying the superpotential $W_u$, 
(1.8). 
By considering a double seesaw mechanism, 
a bilinear form 
$\langle \Phi_u \rangle \langle \Phi_u \rangle$
can directly couple to the up-quark sector $q \, u^c$. 
As seen in Eq.(2.11) and Fig.2 in the next section, 
we would like to emphasize that such the double seesaw 
mechanism becomes possible only when we consider that
${\bf 10}'_\alpha$ is a triplet of the O(3) family symmetry.
Hereafter, we will denote $\Phi_u$ as $\hat{Y}_u$.
Thereby, the $\langle \hat{Y}_u \rangle$-$\langle Y_d \rangle$
correspondence becomes more natural, i.e.
$$\langle \hat{Y_u} \rangle = k'_u \langle \Phi_e \rangle
({\bf 1} + a_u X) \langle \Phi_e \rangle \ \ 
\leftrightarrow \ \ 
\langle Y_d \rangle = k'_d \langle \Phi_e \rangle
({\bf 1} + a_d X) \langle \Phi_e \rangle ,
\eqno(1.15)
$$
compared with Eqs.(1.5) and (1.6).

(ii) {\it New model for the factor $({\bf 1}+a_f X)$}:  
So far, it has been considered that the factors $({\bf 1}+a_f X)$ 
in the VEV relations (1.5) and (1.6) originate in 
VEVs $\langle E \rangle=v_E {\bf 1}$ and 
$\langle S \rangle = v_X X$ of new scalars $E$ and $S$.
In the present model, we consider that the factors originate 
in an S$_3$ invariant coefficients for $\Phi_e \Phi_e$ 
as we discuss in Sec.3.
In the previous SU(5) compatible yukawaon model, since we 
considered that $\Phi_e (E+ a_f S) \Phi_e$ are cubic forms 
of fields, a complicated mechanism was required to 
obtain the VEV relations (1.5) and (1.6).
In the present model, since the factors $({\bf 1}+a_f X)$ 
are merely numerical coefficients, we can present 
the relations (1.5) and (1.6) with a simple 
mechanism.
This change is practically important to build a model without 
a cut off scale $\Lambda$.
In Sec.3, we will assume that the fundamental yukawaon 
$\Phi_e$ obeys a transformation of a permutation symmetry
S$_3$.

Such a modification in a yukawaon model causes considerable change
from previous yukawaon models.
Especially, in contrast to past yukawaon models which are based on 
an effective theory with a cut off $\Lambda$ and with a single family
symmetry, the present yukawaon model somewhat becomes complicated.
However, we consider that it is important to investigate a possibility 
that family symmetry effects are visible, even we 
pay the cost of complicated forms of the superpotential.

\vspace{5mm}

{\large\bf 2. Would-be Yukawa interactions}

Let us consider a superpotential form for would-be Yukawa interactions 
straightforwardly as
$$
W_Y =  \frac{y_u}{\Lambda} {\bf 10}_i Y_{(10,10)}^{ij} {\bf 10}_j {\bf 5}_H 
+ \frac{y_{d,e}}{\Lambda}\bar{{\bf 5}}_i  Y_{(5,10)}^{ij} {\bf 10}_j \bar{\bf 5}_H 
+ \frac{y_\nu}{\Lambda} \bar{{\bf 5}}_i  Y_{(5,1)}^{ij}{\bf 1}_j {\bf 5}_H 
+\lambda_R {\bf 1}_i Y_{(1,1)}^{ij} {\bf 1}_j ,
\eqno(2.1)
$$
where $\bar{{\bf 5}}+{\bf 10}+{\bf 1}$ are quark and lepton fields and
${\bf 5}_H $ and $\bar{\bf 5}_H$ correspond to the conventional 
two Higgs doublets $H_u$ and $H_d$, respectively.  
In the would-be Yukawa interactions (2.1), the charged lepton 
yukawaon $Y_e$ is identical with the down-quark yukawaon $Y_d$, 
i.e. $Y_e=Y_d=Y_{(5,10)}$. 
In the yukawaon model, the yukawaon $Y_e$ has to be different from $Y_d$.
A splitting mechanism between $Y_e$ and $Y_d$ is needed.
Therefore, first, let us give a brief review a $Y_e$-$Y_d$ splitting 
mechanism which has been proposed in the previous SU(5)-compatible
yukawaon model \cite{SU5_1110} with one family symmetry U(3).
We introduce vector-like 
${\bf 5}^{\prime\, i}$ and $\bar{\bf 5}'_i$ fields in addition to 
the fields given in Eq.(1.2). 
For convenience, we denote one ${\bf 5}$ and 
two $\bar{\bf 5}$ as 
$$
\bar{\bf 5}_i = (D_i^c, \ell_i), \ \ \ 
\bar{\bf 5}^{\prime\prime}_i = (d_i^c, L_i), \ \ \ 
{\bf 5}^{\prime\prime\, i} =( \bar{D}^{c\, i}, \bar{L}^{i}), 
\eqno(2.2)
$$
where $d^c$, $D^c$ and $\bar{D}^c$ are SU(2)$_L$ singlet 
down-quarks with electric charges $+1/3$, $+1/3$
and $-1/3$, respectively, and $\ell$, $L$ and $\bar{L}$ 
are SU(2)$_L$ lepton doublets.
In order to realize that the fields $(D^c, \bar{D}^c)$, 
and $(L, \bar{L})$ become massive and decouple from 
the present model, we assume the following interactions
$$
\lambda_D \bar{\bf 5}_i^A  (\Sigma_3)_A^B {\bf 5}^{\prime\prime\, i}_B 
+ \lambda_L \bar{\bf 5}^{\prime\prime A}_i (\Sigma_2)_B^C  
{\bf 5}^{\prime\prime\, i}_B ,
\eqno(2.3)
$$
where $A, B$ are indices of SU(5), and  
SU(5) ${\bf 24}+{\bf 1}$ fields $\Sigma_2$ and $\Sigma_3$ 
take VEV forms 
$$
\begin{array}{l}
\langle \Sigma_2 \rangle = v_{2}\, {\rm diag}(0,0,0,1,1) , \\ 
\langle \Sigma_3 \rangle = v_{3}\, {\rm diag}(1,1,1, 0, 0) .
\end{array}  
\eqno(2.4)
$$
Therefore, Eq.(2.3) leads to mass terms
$$
\lambda_D v_{3}\bar{D}^{c\, i} D_i^c  + \lambda_L v_{2} \bar{L}^i L_i .
\eqno(2.5)
$$
We consider that the VEVs of $\Sigma_2$ and $\Sigma_3$ are 
of the order of $\Lambda_{GUT}$.
As we seen in Eq.(2.3), $R$ charges of $\Sigma_2$ and $\Sigma_3$ 
(and also ${\bf 5}$ and ${\bf 5}'$) have to be different from 
each other:
$$
R(\Sigma_3)-R(\Sigma_2) = R(\bar{\bf 5}_i^{\prime\prime})
- R(\bar{\bf 5}_i ) =R(Y_e) -R(Y_d) .
\eqno(2.6)
$$
If we accept such fields with the VEV forms (2.4), we may understand
the doublet-triplet splitting of the Higgs fields ${\bf 5}_H$ and
$\bar{\bf 5}_H$ by a similar mechanism $\lambda_H \bar{\bf 5}_H
\Sigma_3 {\bf 5}$.
The doublet-triplet splitting mechanism has been already proposed
in the framework of an SO(10) GUT scenario \cite{DT_splitting}.
Therefore, the VEV forms (2.4) will also be understood from 
a GUT scenario based on a higher gauge group and/or on 
extra-dimensions.
For the time being, we do not ask the origin of the VEV forms (2.4). 
This is still an open question.

For such the fields $\bar{\bf 5}_i$ and $\bar{\bf 5}_i^{\prime\prime}$, 
the would-be Yukawa interactions are given by the following superpotential:
$$
W_{Ye,d} = y_{e} \bar{\bf 5}_i Y_e^{i\alpha} {\bf 5}_\alpha^{\prime} 
+  y_{d} \bar{\bf 5}_i^{\prime\prime} Y_d^{i\alpha} {\bf 5}_\alpha^{\prime} 
+M_{5} {\bf 5}_\alpha^{\prime} \bar{\bf 5}^{\prime}_\alpha  
+ y_{5} \bar{\bf 5}^{\prime}_\alpha {\bf 10}_\alpha 
\bar{\bf 5}_H .
\eqno(2.7)
$$
[Here and hereafter, for convenience, we sometime denote 
a field $A_\alpha$ as $A^\alpha$ although those are 
identical because $\alpha$ is an index of O(3). ]
Then, we can obtain the effective Yukawa interaction 
$$
W_{Ye,d}^{eff} = 
\frac{y_e y_{5}}{\bar{M}_{5}} \bar{\bf 5}_i Y_{e}^{i\alpha} 
{\bf 10}_\alpha \bar{\bf 5}_H
+ \frac{y_d y_{5}}{\bar{M}_{5}} \bar{\bf 5}_i^{\prime\prime}
 Y_{d}^{i\alpha} {\bf 10}_\alpha \bar{\bf 5}_H ,
 \eqno(2.8)
$$
where $\bar{M}_5$ is given by 
$$
\bar{M}_5 \simeq \sqrt{ (M_5)^2 + y_{e,d}^2 \langle Y_{e,d} \rangle^2 } ,
\eqno(2.9)
$$
under the approximation $y_5^2 \langle H_d \rangle^2
\ll (M_5)^2$ and in the diagonal basis of 
$\langle Y_{e,d} \rangle$. 
The relation (2.9) has been obtain from the diagonalization 
of mass matrix for $(\bar{\bf 5}\ \bar{\bf 5}', {\bf 10}, 
\bar{\bf 5}^{\prime\prime}, {\bf 5}^{\prime\prime})$
$$
\left( 
\begin{array}{cccc}
0 & 0 & 0 & y_{e,d} \langle Y_{e,d}\rangle \\
0 & 0 & y_5 v_{Hd} & 0 \\
0 &  v_{Hd} & 0 & M_5 \\
y_{e,d} \langle Y_{e,d}\rangle & 0 & M_5 & 0
\end{array} \right) .
\eqno(2.10)
$$
Note that we can use the relation (2.9) even for the case 
$y_{e,d} \langle Y_{e,d}\rangle \sim M_5$.
(For the mass generation mechanism of the charged leptons, 
see Fig.1.)

\begin{figure}[t!]
\begin{center}
\begin{picture}(270,130)
\put(50,50){\thicklines\vector(1,0){25}}
\put(75,50){\thicklines\line(1,0){25}}
\put(100,50){\circle*{4}}
\put(75,30){\large\bf $\overline{\bf 5}_i$}
\put(100,73){\thicklines\vector(0,-1){8}}
\put(100,85){\thicklines\line(0,-1){10}}
\put(100,60){\thicklines\line(0,-1){8}}
\put(95,80){\thicklines\line(1,1){10}}
\put(105,80){\thicklines\line(-1,1){10}}
\put(90,95){\large $\langle Y_e^{i\alpha} \rangle$}
\put(150,50){\thicklines\vector(-1,0){25}}
\put(125,50){\thicklines\line(-1,0){25}}
\put(150,50){\circle*{4}}
\put(125,30){\large ${\bf 5}_\alpha^{\prime}$}
\put(145,55){\large $M_{5}$}
\put(150,50){\thicklines\vector(1,0){25}}
\put(175,50){\thicklines\line(1,0){25}}
\put(200,50){\circle*{4}}
\put(175,30){\large\bf $\overline{\bf 5}^{\prime}_\alpha$}
\put(200,73){\thicklines\vector(0,-1){8}}
\put(200,85){\thicklines\line(0,-1){10}}
\put(200,60){\thicklines\line(0,-1){8}}
\put(195,80){\thicklines\line(1,1){10}}
\put(205,80){\thicklines\line(-1,1){10}}
\put(190,95){\large $\langle \overline{\bf 5}_H \rangle$}
\put(250,50){\thicklines\vector(-1,0){25}}
\put(225,50){\thicklines\line(-1,0){25}}
\put(225,30){\large\bf ${\bf 10}_\alpha$}
\end{picture}
\end{center}
\vspace*{-17mm}

  \caption{Mass generation mechanism for the charged leptons.
}
  \label{M_e}
\end{figure}
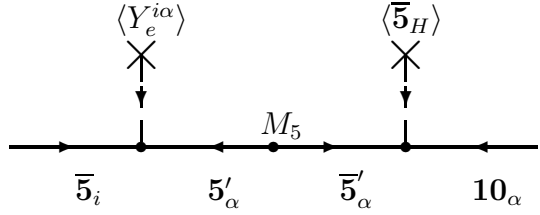

On the other hand, for the up-quark sector, we somewhat
change our model from the previous model (1.5).
As we discuss in the next section, in the yukawaon model, 
the VEV matrix of $Y_u$ is given by a bilinear form 
$\langle Y_u \rangle = k_u \langle \Phi_u \rangle 
\langle \Phi_u \rangle$, and $\langle Y_d \rangle$
takes the same structure as $\langle \Phi_u \rangle$
except for values of the parameters $a_{u}$ and $a_{u}$. 
Therefore, in this paper, we denote $\Phi_u$ in the previous paper 
as $\hat{Y}_u$, and we propose a model without $Y_u$ in the 
old model:
$$
W_{Yu} = y_{u} {\bf 10}_\alpha \hat{Y}_u^{\alpha \beta} 
\overline{\bf 10}_\beta^{\prime} 
+M_{10} \overline{\bf 10}_\alpha^{\prime}  
{\bf 10}^{\prime}_\alpha
+ y_{10} {\bf 10}^{\prime}_\alpha 
{\bf 10}^{\prime}_\alpha {\bf 5}_H .
\eqno(2.11)
$$
Note that the Higgs field ${\bf 5}_H$ couples not to 
${\bf 10}^\prime {\bf 10}$, but to ${\bf 10}^\prime {\bf 10}'$,
differently from Eq.(1.8).
Therefore, the effective interaction is given by a double seesaw
form
$$
W_{Yu}^{eff} = \frac{y_u^2 y_{10}}{(\overline{M}_{10})^2} 
{\bf 10}_\alpha 
\hat{Y}_{u}^{\alpha \gamma} \hat{Y}_{u}^{\gamma \beta}
{\bf 10}_\beta {\bf 5}_H ,
\eqno(2.12)
$$
under the approximation $\langle H_u \rangle \ll M_{10}$.
We again would like to emphasize that the double seesaw form 
(2.12) is possible only when we consider the third term
in Eq.(2.11), i.e. only when ${\bf 10}'_\alpha$ is a triplet of O(3) 
family symmetry.
In Eq.(2.10), $\overline{M}_{10}^{\alpha\beta}$ is given by 
$$
\overline{M}_{10} \simeq M_{10} 
\eqno(2.13)
$$
from the diagonalization of the mass matrix for the fields 
$({\bf 10}, {\bf 10}', \bar{\bf 10}' )$
$$
\left(
\begin{array}{ccc}
0 & 0 & y_u \langle Y_u \rangle \\
0 & y_{10} v_{Hu} & M_{10} \\
y_u \langle Y_u \rangle & M_{10} & 0
\end{array} \right) .
\eqno(2.14)
$$
We can use the relation (2.13) even for a case of 
$y_u \langle Y_u \rangle \sim M_{10}$, 
although $\overline{M}_{10}$ in the previous SU(5) compatible 
model has been highly dependent of the value of
$\langle \hat{Y}_{u} \rangle$ \cite{SU5_1110}.
This is because the Higgs field ${\bf 5}_H$ couples to
${\bf 10}'_\alpha {\bf 10}'_\alpha$ in the present model, 
not to ${\bf 10}^{\prime\, i} {\bf 10}_i$ as in the previous
model.

\begin{figure}[t!]

\begin{center}
\begin{picture}(370,130)
\put(50,50){\thicklines\vector(1,0){25}}
\put(75,50){\thicklines\line(1,0){25}}
\put(100,50){\circle*{4}}
\put(75,30){\large\bf ${\bf 10}_\alpha$}
\put(100,73){\thicklines\vector(0,-1){8}}
\put(100,85){\thicklines\line(0,-1){10}}
\put(100,60){\thicklines\line(0,-1){8}}
\put(95,80){\thicklines\line(1,1){10}}
\put(105,80){\thicklines\line(-1,1){10}}
\put(90,95){\large $\langle \hat{Y}_u^{\alpha\gamma} \rangle$}
\put(150,50){\thicklines\vector(-1,0){25}}
\put(125,50){\thicklines\line(-1,0){25}}
\put(150,50){\circle*{4}}
\put(125,30){\large $\overline{\bf 10}_\gamma^{\prime}$}
\put(145,57){\large $M_{10}$}
\put(150,50){\thicklines\vector(1,0){25}}
\put(175,50){\thicklines\line(1,0){25}}
\put(200,50){\circle*{4}}
\put(175,30){\large\bf ${\bf 10}^{\prime}_\gamma$}
\put(200,73){\thicklines\vector(0,-1){8}}
\put(200,85){\thicklines\line(0,-1){10}}
\put(200,60){\thicklines\line(0,-1){8}}
\put(195,80){\thicklines\line(1,1){10}}
\put(205,80){\thicklines\line(-1,1){10}}
\put(190,95){\large $\langle {\bf 5}_H \rangle$}
\put(250,50){\thicklines\vector(-1,0){25}}
\put(225,50){\thicklines\line(-1,0){25}}
\put(225,30){\large\bf ${\bf 10}'_\gamma$}
\put(250,50){\circle*{4}}
\put(245,57){\large $M_{10}$}
\put(250,50){\thicklines\vector(1,0){25}}
\put(275,50){\thicklines\line(1,0){25}}
\put(300,50){\circle*{4}}
\put(275,30){\large\bf ${\bf 10}'_\gamma$}
\put(300,73){\thicklines\vector(0,-1){8}}
\put(300,85){\thicklines\line(0,-1){10}}
\put(300,60){\thicklines\line(0,-1){8}}
\put(295,80){\thicklines\line(1,1){10}}
\put(305,80){\thicklines\line(-1,1){10}}
\put(290,95){\large $\langle \hat{Y}_u^{\gamma\beta} \rangle$}
\put(350,50){\thicklines\vector(-1,0){25}}
\put(325,50){\thicklines\line(-1,0){25}}
\put(325,30){\large $\overline{\bf 10}_\beta$}
%
%
\end{picture}
\end{center}
\vspace*{-17mm}

  \caption{Mass generation mechanism for the up-quarks.
}
  \label{M_u}
\end{figure}
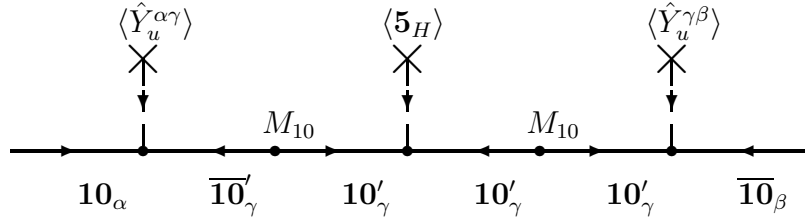

Thus, SU(5) non-singlet fields which can contribute
to the evolutions of the gauge coupling constants of
SU(3)$_c \times$SU(2)$_L \times$U(1)$_Y$ below 
$\mu < \Lambda_{GUT}$ are only
$$
({\bf 5}_\alpha^{\prime} +
\bar{\bf 5}^{\prime}_\alpha)
+(\overline{\bf 10}_\alpha^{\prime} + 
{\bf 10}^{\prime}_\alpha)
 .
\eqno(2.15)
$$
in addition to the standard $\bar{\bf 5}_i + {\bf 10}_\alpha$.
The mass parameters $M_5$ and $M_{10}$ are free parameters
in the superpotential.
We will consider $M_5 \ll M_{10}$ in the next section. 
On the other hand, the fields 
$(D_i^c, L_i)$ and $(\bar{D}^{ci}, \bar{L}^i)$
given in Eq.(2.1) 
cannot contribute to the evolutions of gauge coupling
constants of SU(3)$_c \times$SU(2)$_L \times$U(1)$_Y$, 
because those particles have 
masses of the order of $\Lambda_{GUT}$.

  
Next, we discuss a seesaw-type mass matrix for neutrinos. 
Differently from the previous SU(5) compatible yukawaon model \cite{SU5_1110},
we introduce an SU(5) singlet field ${\bf 1}_\alpha$ instead of 
${\bf 1}_i$ in the previous model.
The neutrino Dirac mass term $m_D$ is obtained from the following
superpotential 
$$
W_{Y\nu} = y_e \bar{\bf 5}_i Y_e^{i\alpha} {\bf 5}_\alpha^{\prime} 
+M_{5} {\bf 5}_\alpha^{\prime} \bar{\bf 5}^{\prime}_\alpha 
+ y_{1} \bar{\bf 5}^{\prime}_\alpha {\bf 1}_\alpha {\bf 5}_H ,
\eqno(2.16)
$$
where only the third term is a new term and the first and second 
terms have already given in Eq.(2.7).
The superpotential (2.16) leads to the effective interaction
$$
W_{Y\nu}^{eff} = \frac{y_e y_{1}}{M_{5}} \bar{\bf 5}_i Y_{e}^{i\alpha} 
{\bf 1}_\alpha {\bf 5}_H .
\eqno(2.17)
$$
Note that the neutrino Dirac mass matrix has the same structure
as the charged lepton mass matrix. 
On the other hand, the right-handed Majorana neutrino mass
matrix $M_R$ is obtained from the superpotential term
$$
W_R = \lambda_R {\bf 1}_\alpha Y_R^{\alpha\beta} {\bf 1}_\beta .
\eqno(2.18)
$$
Therefore, 
we can obtain a seesaw-type neutrino mass matrix
$$
M_\nu = \frac{y_e^2 y_1^2}{\lambda_R} \left( 
\frac{v_{Hu}}{M_5} \right)^2 
\langle Y_e \rangle \langle Y_R \rangle^{-1} 
\langle Y_e \rangle .
\eqno(2.19)
$$
From Eq.(2.8), we can rewritten Eq.(2.19) as
$$
M_\nu = \frac{y_1^2 \tan^2 \beta}{\lambda_R y_5^2}  \, 
M_e \, \langle Y_R \rangle^{-1} M_e ,
\eqno(2.20)
$$
where $ \tan\beta =\langle H_u^0 \rangle/\langle H_d^0 \rangle$.
By taking $m_\tau \simeq 1.777$ GeV, $(M_{\nu })_{33}
\sim m_{\nu 3} \simeq \sqrt{\Delta m_{atm}^2} \simeq 0.049$ eV 
\cite{PDG10}) and $\tan\beta \simeq 10$,
we can estimate the value of $\langle Y_R \rangle$ as
$$
\langle Y_R \rangle  \simeq \lambda_R (y_5/y_1)^2 \times 6.4
\times 10^{12} \ {\rm GeV}.
\eqno(2.21)
$$
However, this does not mean that the value of $\Lambda_{O3}$ is 
of the order of $\langle Y_R^{\alpha\beta} \rangle
\sim 10^{12-13}$ GeV.
The value of $\Lambda_{O3}$ is determined by the largest one of 
all O(3)-non-singlet scalars.
We can assert only $\Lambda_{O3} \ge \langle Y_R^{\alpha\beta} \rangle$.

\vspace{5mm}

{\large\bf 3. Yukawaon sector}

Priori to discussing VEV relations among yukawaons, we discuss a new idea 
about the factors $({\bf 1}+a_f X)$ in Eqs.(1.5) and (1.6). 
The factors play an essential role in giving successful results 
in the phenomenological yukawaon model.
In the past yukawaon model, it has been considered that the factors 
$({\bf 1}+a_f X)$ are originated in 
VEVs of scalars $E$ and $S$, so that $\Phi_e (E+ a_f S)\Phi_e$ 
were cubic forms of fields. 
As a result, in order to build a model without $\Lambda$ and 
in order to obtain the VEV forms given in Eq.(1.7), very complicated 
mechanism was required.  
In the present model, since the factors $({\bf 1}+a_f X)$ 
are merely numerical coefficients, 
$\Phi_e ({\bf 1}+a_f X)\Phi_e$ is a bilinear form of the fields (not 
a cubic form of fields). 

When we denote a doublet $(\psi_\pi, \psi_\eta)$ and 
a singlet $\psi_\sigma$ in a permutation symmetry \cite{S3} S$_3$ as
$$
\left( 
\begin{array}{c}
\psi_\pi \\
\psi_\eta 
\end{array} \right) = \left(
\begin{array}{c}
\frac{1}{\sqrt2} (\psi_1 -\psi_2) \\
\frac{1}{\sqrt6} (\psi_1 +\psi_2 -2\psi_3)
\end{array} \right) ,
\eqno(3.1)
$$
$$
\psi_\sigma =\frac{1}{\sqrt3}(\psi_1+\psi_2+\psi_3) ,
\eqno(3.2)
$$
the field $\psi = (\psi_1,\psi_2,\psi_3)$ is represented as
$$
\psi \equiv \left( 
\begin{array}{c}
\psi_1 \\
\psi_2 \\
\psi_3
\end{array} \right) = \left(
\begin{array}{ccc}
\frac{1}{\sqrt2} & \frac{1}{\sqrt6} & \frac{1}{\sqrt3} \\
-\frac{1}{\sqrt2} & \frac{1}{\sqrt6} & \frac{1}{\sqrt3} \\
0 & -\frac{2}{\sqrt6} & \frac{1}{\sqrt3} \\
\end{array} \right) \left(
\begin{array}{c}
\psi_\pi \\
\psi_\eta \\
\psi_\sigma
\end{array} \right) 
\equiv U_0 \left(
\begin{array}{c}
\psi_\pi \\
\psi_\eta \\
\psi_\sigma
\end{array} \right) .
\eqno(3.3)
$$
A bilinear form $\psi\psi$ is invariant under the S$_3$ 
symmetry only when $\psi_a \xi_{ab} \psi_b$
is given by the form
$$
\psi_a \xi_{ab} \psi_b =
\psi_a ({\bf 1}+a X)_{ab} \psi_b ,
\eqno(3.4)
$$
where $a$ is a free parameter, and 
${\bf 1}$ and $X$ are defined by Eq.(1.7). 
The matrix $X$ is diagonalized by $U_0$ as
$$
U_0^T X U_0 = \left(
\begin{array}{ccc}
0 & 0 & 0 \\
0 & 0 & 0 \\
0 & 0 & 1
\end{array} \right) .
\eqno(3.5)
$$ 

Therefore, in the present model, we assume that the fundamental 
yukawaon $\Phi_e$ transforms as $\psi$ defined by Eq.(3.3) 
under the S$_3$ symmetry, i.e. as $\Phi_e^{a \alpha}$.
We assume the following  S$_3$ invariant superpotential terms
$$
W_e = \left( \mu_e Y_e^{i\alpha}
+\lambda_e \bar{E}^{i\gamma} \hat{Y}_e^{\gamma \alpha} \right)
\Theta^e_{\alpha i} 
+ \left(\mu'_e \hat{Y}_e^{\alpha\beta}  
+\lambda'_e \Phi_e^{T\alpha a} \xi^e_{ab} \Phi_e^{b\beta} \right)
\Theta^{e\prime}_{\beta\alpha} ,
\eqno(3.6)
$$
$$
W_d = \left( \mu_d Y_d^{i\alpha}
+\lambda_d \bar{E}^{i\gamma} \hat{Y}_d^{\gamma \alpha }\right)
 \Theta^d_{\alpha i} 
+ \left[ \lambda'_d (\hat{E}^{\alpha\gamma} \hat{Y}_d^{\gamma\beta}
+ \hat{Y}_d^{\alpha\gamma} \hat{E}^{\gamma\beta}) 
+  \lambda^{\prime\prime}_d \Phi_e^{T\alpha a} \xi^d_{ab} \Phi_e^{b\beta} 
\right] \Theta^{d\prime}_{\beta\alpha}  ,
\eqno(3.7)
$$
$$
W_u = \left[\lambda_u \left( P_u^{\alpha\gamma} \hat{Y}_u^{\gamma\beta} 
+ \hat{Y}_u^{\alpha\gamma} P_u^{\gamma\beta} \right) 
+\lambda'_u \Phi_e^{T\alpha a} \xi^u_{ab} \Phi_e^{b\beta} \right]
\Theta^{u}_{\beta\alpha} ,
\eqno(3.8)
$$
where 
$$
\xi^f_{ab} =({\bf 1}+a_f X)_{ab} .
\eqno(3.9)
$$
In Eqs.(3.6) and (3.7), $Y_e$ and $Y_d$ are connected to
$(\Phi_e^T \xi \Phi_e)$ via two steps. 
The introducing $\hat{Y}_e$ is to connect $Y_R^{\alpha\beta}$ to 
$\hat{Y} _e^{\alpha\gamma} \hat{Y}_u^{\gamma\beta}$ as seen later.
Then, it is required that $\hat{Y} _e$ is distinguished from
$\hat{Y} _d$ by $R$ charges.
Therefore, we have assumed different structures for 
$\hat{Y} _e$ and $\hat{Y} _d$ as given in Eqs.(3.6) and (3.7).
Also, the field $P_u$ has been inserted in Eq.(3.8) in order to
distinguish $\hat{Y}_u$ from $\hat{Y}_e$ and $\hat{Y}_d$ under the 
$R$ charge conservation. 
Since
$$
R(\hat{E}) + R(\hat{Y}_d) = 2R(\Phi_e)=R(\hat{Y}_e) ,
\eqno(3.10)
$$
$$
R(\hat{Y}_u) = R(\hat{Y}_e) -R(P_u) = R(\hat{Y}_d)
+R(\hat{E})-R(P_u) ,
\eqno(3.11)
$$
we can distinguish $\hat{Y} _d$ from $\hat{Y} _e$ and
$\hat{Y} _u$ when $R(\hat{E}) \neq 0$ and 
$R(\hat{E}) \neq R(P_u)$, respectively.

The values of $a_f$ in Eq.(3.9) are purely phenomenological
parameters. 
At present, there is no reason that we take $a_e=0$.
However, we think that the VEV matrix $\langle \Phi_e 
\rangle$ is a fundamental VEV matrix in the model, 
so that it is likely that the value $a_e$ in 
$\langle \hat{Y}_e \rangle$ takes a specific value 
$a_e=0$. 
However, the true reason is a future task to us.

By using SUSY vacuum conditions 
$\partial W/\partial \Theta_A =0$
($\Theta_A= \Theta^e, \Theta^{e\prime}, \Theta^d, 
 \Theta^{d\prime},
\Theta^u$) for the superpotential terms (3.6)-(3.8), 
and by assuming that our vacuum takes 
$\langle \Theta_A \rangle = 0$, 
we obtain the following VEV relations:
$$
\langle Y_e^{i\alpha} \rangle = -\frac{\lambda_e}{\mu_e}
\langle \bar{E}^{i \gamma}\rangle 
\langle \hat{Y}_e^{\gamma\alpha} \rangle 
=  \frac{\lambda_e \lambda'_e}{\mu_e \mu'_e}
\langle \bar{E}^{i \gamma} \rangle 
\langle \Phi_e^{T\alpha a}\rangle \xi^e_{ab} 
\langle \Phi_e^{b\beta}\rangle ,
\eqno(3.12)
$$
$$
\langle Y_d^{i\alpha} \rangle = -\frac{\lambda_d}{\mu_d}
\langle \bar{E}^{i \gamma}\rangle
\langle \hat{Y}_d^{\gamma\alpha} \rangle 
=  \frac{\lambda_d \lambda^{\prime\prime}_d}{2\mu_d \lambda'_d}
\langle \bar{E}^{i \gamma}\rangle
\langle (\hat{E}^{-1})^{\gamma\delta} \rangle 
\langle \Phi_e^{T\delta a}\rangle \xi^d_{ab} 
\langle \Phi_e^{b\alpha}\rangle ,
\eqno(3.13)
$$
$$
\langle P_u^{\alpha\gamma} \rangle\langle \hat{Y}_u^{\gamma\beta} \rangle 
+\langle \hat{Y}_u^{\alpha\gamma} \rangle \langle P_u^{\gamma\beta} \rangle
= -\frac{\lambda'_d}{\lambda_d}  
\langle \Phi_e^{T\gamma a}\rangle \xi^u_{ab} 
\langle \Phi_e^{b\beta}\rangle ,
\eqno(3.14)
$$
where we assume that the VEV forms of $\langle \bar{E} \rangle$ and 
$\langle P_u^{\alpha\gamma}\rangle $ are given by 
$$
\frac{1}{\bar{v}_E} \langle \bar{E} \rangle_e = 
\frac{1}{\hat{v}_E} \langle \hat{E} \rangle_e = 
\left( 
\begin{array}{ccc}
1 & 0 & 0 \\
0 & 1 & 0 \\
0 & 0 & 1
\end{array} \right) , \ \ \ \ 
\frac{1}{v_{Pu}} \langle P_u \rangle_u =  \left( 
\begin{array}{ccc}
1 & 0 & 0 \\
0 & -1 & 0 \\
0 & 0 & 1
\end{array} \right) .
\eqno(3.15)
$$
Here, the expression $\langle A \rangle_f$ ($f=e,u$) denotes 
a form of the VEV matrix $\langle A \rangle$ in a basis
(we call it $f$-basis) 
in which the mass matrix $M_f$ is diagonal.  
Note that almost VEV forms are represented with simple forms 
in the $e$-basis, 
while only $\langle P_u \rangle$ takes a simple form (3.15)
in the $u$-basis. 
Therefore, the assumption for the form (3.15) is somewhat 
strange. 
Such the form (3.15) was introduce \cite{O3_09PLB} in order to change the 
sign $(+,-,+)$ of the eigenvalues of $\langle \Phi_u \rangle$
(i.e. $\langle \hat{Y}_u \rangle$ in the present model) 
to the positive values $(+,+,+)$.
We needs the field $P_u$ 
in order to obtain the successful fitting for the observed
neutrino mixing and up-quark mass ratios
as we discuss in the next section. 
 
Here and hereafter, we denote fields whose VEV values
are zeros as $\Theta_A$ ($A=e, u, \cdots$).
Therefore, we can obtain meaningful VEV relations from
SUSY vacuum conditions $\partial W/\partial \Theta_A=0$, 
while we cannot obtain any relations 
from other conditions (e.g. $\partial W/\partial Y_f=0$) 
because the relations always include $\langle \Theta_A \rangle$.
For the time being, we assume that the supersymmetry breaking 
is induced by a gauge mediation mechanism (not including family 
gauge symmetries), 
so that our VEV relations among yukawaons are still valid 
even after
the SUSY was broken in the quark and lepton sectors.

Finally, we comment on the VEV forms of $\bar{E}$, 
$\hat{E}$ and $P_u$ which were assumed as in Eq.(3.15).
We cannot directly give the forms (3.15),
but we can give the relations 
$$
\langle \hat{E} \rangle = \hat{v}_E
{\bf 1} , \ \ \ 
\langle {E} \rangle \langle \bar{E} \rangle = v_E \bar{v}_E
{\bf 1} , \ \ \ 
\langle {P}_u \rangle^2  = v_P^2 {\bf 1} , 
\eqno(3.16)
$$
by introducing a new field $E_{i\alpha}$ and  
by assuming the following superpotential
$$
W_{E,P} = \lambda_1 {\rm Tr}[E \bar{E} \hat{E} ]
+\lambda_2 {\rm Tr}[E \bar{E} ] {\rm Tr}[\hat{E}]
+\lambda_3 {\rm Tr}[P_u P_u P_u] + \lambda_4 
{\rm Tr}[E \bar{E}] {\rm Tr}[P_u ] ,
\eqno(3.17)
$$
where we have assumed
$$
R(\hat{E})=R(P_u)= \frac{2}{3} , \ \ \ \ 
R(E) +R(\bar{E}) = \frac{4}{3} .
\eqno(3.18)
$$
The SUSY vacuum conditions $\partial W/\partial \hat{E} =0$ 
and $\partial W/\partial P_u =0$ can give
$$
\frac{\partial W}{\partial \hat{E}} =
\lambda_1 E \bar{E} + \lambda_2 {\rm Tr}[E \bar{E} ] {\bf 1}=0,
\eqno(3.19)
$$
$$
\frac{\partial W}{\partial P_u} = 3\lambda_3 P_u P_u 
+ \lambda_4 {\rm Tr}[E \bar{E}] {\bf 1} = 0,
\eqno(3.20)
$$
which lead to the relations 
$\langle {E} \rangle \langle \bar{E} \rangle \propto {\bf 1}$
and $\langle {P}_u \rangle^2 \propto {\bf 1}$, respectively.
The remaining conditions  $\partial W/\partial E_f =0$ 
and $\partial W/\partial \bar{E}^f =0$ can be satisfied for the
case $\langle \hat{E}_f \rangle = v_E {\bf 1}$.
We consider that the forms (3.15) are specific solutions of (3.16).

%


Next, we discuss a possible form of $\langle Y_R \rangle$.
In the previous O$_3$ yukawaon model \cite{O3_09PLB}, 
the form $\langle {Y}_R \rangle$ has been given by
$$
\langle Y_R \rangle_e = k_R \left[ \langle \Phi_u \rangle_e
\langle P_u \rangle_e \langle Y_e \rangle_e +
\langle Y_e \rangle_e \langle P_u \rangle_e 
\langle \Phi_u \rangle_e  + \xi_\nu (
\langle \Phi_u \rangle_e \langle Y_e \rangle_e 
\langle P_u \rangle_e + \langle P_u \rangle_e
\langle Y_e \rangle_e \langle \Phi_u \rangle_e )
\right] , 
\eqno(3.21)
$$
where $\langle P_u \rangle_u$ is given by Eq.(3.15).
In contrast to Eq.(3.21), in the present model, 
$\langle Y_R \rangle$ is derived from the following superpotential
$$
W_R = \left\{ \mu_R Y_R^{\alpha\beta} +
\lambda'_R \left[ \hat{Y}_u^{\alpha\gamma} \hat{Y}_e^{\gamma\beta} 
+ \hat{Y}_e^{\alpha\gamma} \hat{Y}_u^{\gamma\beta} 
+\xi_\nu \left( {\rm Tr}(\hat{Y}_u) \hat{Y}_e^{\alpha\beta} 
+ {\rm Tr}(\hat{Y}_e) \hat{Y}_u^{\alpha\beta} \right) \right] \right\}
\Theta^R_{\beta\alpha} ,
\eqno(3.22)
$$
without $P_u$.
Instead, $P_u$ has been inserted in $W_u$ as given in Eq.(3.8). 


We notice that, in Eq.(3.22), the $\xi_\nu$ term has been changed
from Eq.(3.21) in the O(3) model.
Nevertheless, we can again obtain reasonable value of the 
neutrino mixing parameters by fitting the parameters
$a_u$ and $\xi_\nu$:
By using the input value $a_u=-1.78$, we can give
reasonable up-quark mass ratios
$$
\sqrt{\frac{ {m_u}}{{m_c}}} = 0.04389, \ \ \ \   
\sqrt{\frac{ {m_c}}{m_t}} = 0.05564,
\eqno(3.23)
$$
which are in good agreement with 
the observed values at $\mu=m_Z$ \cite{q-mass} 
$\sqrt{ {m_u}/{m_c}} = 0.045^{+0.013}_{-0.010}$ and  
$\sqrt{ {m_c}/{m_t}} = 0.060 \pm 0.005$.
Then, the predicted neutrino oscillation parameters are given 
in Table 1.
The results are in favor of the observed values except for 
that the value of $\sin^2 2\theta_{13}$ is too small. 
For this problem in $\sin^2 2\theta_{13}$, we may improve 
the present model by taking some other small effects 
into consideration.

\begin{table}
\begin{center}
\begin{tabular}{cccc} \hline
$\xi_\nu$ & $\tan^2 \theta_{solar}$ & $\sin^2 2\theta_{atm}$
& $\sin^2 2\theta_{13}$ \\ \hline
$0$    & $0.6995$  & $0.9872$ & $0.00068$ \\
$0.0004$ & $0.4881$ & $0.9880$ & $0.00072$ \\
$0.0005$ & $0.4477$ & $0.9882$ & $0.00073$ \\
$0.0006$ & $0.4112$ & $0.9884$ & $0.00074$ \\
\hline
\end{tabular}  
\end{center}
\begin{quotation}
\caption{
$\xi_\nu$ dependence of the neutrino mixing parameters.
The value of $a_u$ is taken as $a_u=-1.78$
which can give reasonable up-quark mass ratios.
}
\end{quotation}
\end{table}




\begin{table}
\begin{center}
\begin{tabular}{|c|ccccccccccc|} \hline
   & $\bar{\bf 5}_i$ & ${\bf 10}_\alpha$ & ${\bf 1}_\alpha$ & 
$\bar{\bf 5}^{\prime\prime}_i$  & ${\bf 5}^{\prime\prime\, i}$ & 
$\bar{\bf 5}'_\alpha$ & ${\bf 5}'_\alpha$ & 
${\bf 10}^{\prime}_\alpha$ & $\bar{\bf 10}^{\prime}_\alpha$ &
$\bar{\bf 5}_H$ & ${\bf 5}_H$  \\ \hline
SU(5) & ${\bf 5}^*$ & ${\bf 10}$ & ${\bf 1}$ & 
${\bf 5}^*$ & ${\bf 5}$ & ${\bf 5}^*$ & ${\bf 5}$ & 
${\bf 10}$ & ${\bf 10}^*$ & ${\bf 5}^*$ & ${\bf 5}$  \\
U(3) & ${\bf 3}$ & ${\bf 1}$ & ${\bf 1}$ & 
${\bf 3}$ &  ${\bf 3}^*$ & ${\bf 1}$ & ${\bf 1}$ & 
${\bf 1}$ & ${\bf 1}$ & ${\bf 1}$ & ${\bf 1}$ \\
O(3) & ${\bf 1}$ & ${\bf 3}$ & ${\bf 3}$ & ${\bf 1}$ &
 ${\bf 1}$ & ${\bf 3}$ & ${\bf 3}$ & ${\bf 3}$ & 
${\bf 3}$ & ${\bf 1}$ & ${\bf 1}$  \\
 \hline
\end{tabular}

\begin{tabular}{|ccccccccccc|} \hline
 $\Sigma_3$ & $\Sigma_2$ &   
$Y_e^{i\alpha}$ & $\hat{Y}_e^{\alpha\beta}$ &  $\Phi_e^{a\alpha}$ &
$\Theta^e_{\alpha i} $ & $\Theta^{e \prime}_{\alpha\beta} $ &
$Y_d^{i\alpha}$  & $\Theta^d_{\alpha i}$ & 
$\hat{Y}_d^{\alpha\beta}$  & $\Theta^{d\prime}_{\alpha\beta}$
\\ \hline
${\bf 24}+ {\bf 1}$ & ${\bf 24}+ {\bf 1}$ & 
${\bf 1}$ & ${\bf 1}$ & ${\bf 1}$ & ${\bf 1}$ & ${\bf 1}$ & 
 ${\bf 1}$ & ${\bf 1}$ &  ${\bf 1}$ & ${\bf 1}$ \\
${\bf 1}$ & ${\bf 1}$ &  ${\bf 3}^*$ & ${\bf 1}$ & 
${\bf 1}$ & ${\bf 3}$ & ${\bf 1}$ & ${\bf 3}^*$ & ${\bf 3}$  & 
${\bf 1}$ & ${\bf 1}$ \\
 ${\bf 1}$  & ${\bf 1}$ &  ${\bf 3}$ & 
${\bf 5}+ {\bf 1}$ & ${\bf 3}$ & ${\bf 3}$ & ${\bf 5}+ {\bf 1}$ &
${\bf 3}$ & ${\bf 3}$ & ${\bf 5}+ {\bf 1}$ & ${\bf 5}+ {\bf 1}$ \\
\hline
\end{tabular}

\begin{tabular}{|cccccccccccc|} \hline
 $\bar{E}^{i\alpha}$ & ${E}_{\alpha i}$ & $\hat{E}^{\alpha\beta}$ & 
$\hat{Y}^{\alpha\beta}_u$ & 
$P_u^{\alpha\beta}$ & $\Theta^u_{\alpha\beta}$ & 
$Y_R^{\alpha\beta}$ &
$\Theta^R_{\alpha\beta}$ & $\hat{E}_{\alpha\beta}$ & 
${T}_A^{i \alpha}$ & ${T}_B^{i \alpha}$ & $S_i$ 
\\ \hline
 ${\bf 1}$ & ${\bf 1}$ & ${\bf 1}$ & ${\bf 1}$ & 
 ${\bf 1}$ & ${\bf 1}$ & 
${\bf 1}$ & ${\bf 1}$ & ${\bf 1}$ & ${\bf 1}$ & ${\bf 1}$ 
 & ${\bf 1}$ \\
 ${\bf 3}^*$ &  ${\bf 3}^*$ & ${\bf 3}$ &
${\bf 1}$ & ${\bf 1}$ & ${\bf 1}$ & ${\bf 1}$ & ${\bf 1}$ & 
${\bf 1}$ & ${\bf 3}^*$ & ${\bf 3}^*$ & ${\bf 3}$  \\
${\bf 3}$ & ${\bf 3}$ & ${\bf 5}+ {\bf 1}$ &
${\bf 5}+ {\bf 1}$ & ${\bf 5}+ {\bf 1}$ & 
${\bf 5}+ {\bf 1}$ & ${\bf 5}+ {\bf 1}$ &
${\bf 5}+ {\bf 1}$ & ${\bf 5}+ {\bf 1}$ & 
${\bf 3}$ & ${\bf 3}$ & ${\bf 1}$  \\
\hline
\end{tabular}

\end{center}
\begin{quotation}
\caption{
Fields in the present model and their SU(5)$\times$U(3)$\times$O(3)
assignments. 
}
\end{quotation}
\end{table}


In Table 2, we list assignments of SU(5)$\times$U(3)$\times$O(3)
for all fields in the present model. 
Obviously, the present model is anomaly free in SU(5). 
In Table 2, in order to make the model anomaly free in the U(3) family 
symmetry, we have added new fields $T_A^{i\alpha}$, 
$T_B^{i\alpha}$ and $S_i$, 
because we have a sum of the anomaly coefficients $\sum A=19-14=5$
except for $T^{i\alpha}$ and $S_i$. 
However, for the time being, we do not specify the roles of those 
fields $T_A$, $T_B$ and $S$ in the model.
At least, the sterile neutrino $S_i$ is harmless, because the 
sterile neutrino can couple to the massive field ${\bf 5}^{\prime\prime}$ 
(mass$\sim \Lambda_{GUT}$) as 
${\bf 5}^{\prime\prime\, i} S_i \bar{\bf 5}_H$.
The existence of $T^{i\alpha}_A$ and $T^{i\alpha}_B$ will play a role in fitting the 
Cabibbo-Kobayashi-Maskawa mixing parameters.

In the present model, fields which have the same quantum numbers
of SU(5)$\times$U(3)$\times$O(3) are distinguished from others 
by $R$ charges. 
Since we have still free parameters in the assignments of $R$ charges, 
we do not give explicit numerical assignments in 
Table 2.  
 
Finally, we would like to comment on $R$ parity assignments.
Since we inherit $R$ parity assignments in the standard SUSY model,
$R$ parities of yukawaons $Y_f$ (and also $\Theta_f$, $\Phi_{e,u}$, 
$E$, $\cdots$) are the same as those of Higgs particles 
(i.e. $P_R({\rm fermion})=-1$ and $P_R({\rm scalar})=+1$), 
while $(\bar{\bf 5}^{\prime\prime} + {\bf 5}^{\prime\prime})$,
$(\bar{\bf 5}^{\prime} + {\bf 5}^{\prime})$ and 
$({\bf 10}^{\prime} + \overline{\bf 10}^{\prime})$ 
are assigned to quark and lepton type, i.e. 
$P_R({\rm fermion})=+1$ and $P_R({\rm scalar})=-1$.

\vspace{5mm}

{\large\bf 4. Energy scales}

In the present model, we have introduced three energy scales 
$\Lambda_{GUT}$, $\Lambda_{O3}$ and $\Lambda_{U3}$, which 
break SU(5), O(3) and U(3), respectively.
As seen in Eqs.(3.12)-(3.14), if we take
$\mu_e, \ \mu'_e, \ \mu_d, \ \mu_R \sim \Lambda_{O3}$,
we can take VEV values as
$\langle Y_e^{i\alpha} \rangle, \ \langle Y_d^{i\alpha} \rangle, \ 
\langle \bar{E}^{i\alpha} \rangle, \ \langle E_{i\alpha} \rangle \ 
\sim \ \Lambda_{U3}$, and $\langle \hat{Y}_e^{\alpha\beta} \rangle$,  
$\langle \hat{Y}_d^{\alpha\beta} \rangle$,  
$\langle \hat{Y}_u^{\alpha\beta} \rangle$, 
$\langle {Y}_R^{\alpha\beta} \rangle$, 
$\langle \hat{E}^{\alpha\beta} \rangle$,  
$\langle P_u^{\alpha\beta} \rangle$,  
$\langle \Phi_e^{a \alpha} \rangle \ \sim \Lambda_{O3}$,
so as to be consistent with the relations (3.12)-(3.14) and (3.21).
(The expression $\langle A \rangle \sim \Lambda$ for a field $A$ means that 
the largest component of $\langle A \rangle$ is of the order 
of $\Lambda$.)

However, as seen in Table 2, we have many O(3) non-singlet fields in 
the present model.
If we consider $\Lambda_{O3} < \Lambda_{GUT}$, 
the gauge coupling constant of O(3) will rapidly
blow up before $\mu$ reaches $\Lambda_{GUT}$.
Therefore, we are obliged to consider 
$$
\Lambda_{O3} \sim \Lambda_{GUT} .
\eqno(4.1)
$$ 
When we simply take
$$
\mu_e, \ \mu'_e, \ \mu_d, \ \sim \Lambda_{O3} ,
\eqno(4.2)
$$
we can obtain
$$
\begin{array}{l}
\langle Y_e^{i\alpha} \rangle, \ \langle Y_d^{i\alpha} \rangle, \ 
\langle \bar{E}^{i\alpha} \rangle, \ \langle E_{i\alpha} \rangle \ 
\sim \ \Lambda_{U3} , \\
\langle \hat{Y}_e^{\alpha\beta} \rangle, \ 
\langle \hat{Y}_d^{\alpha\beta} \rangle, \ 
\langle \hat{Y}_u^{\alpha\beta} \rangle, \ 
\langle \hat{E}^{\alpha\beta} \rangle, \ 
\langle P_u^{\alpha\beta} \rangle, \   
\langle \Phi_e^{a \alpha} \rangle \ \sim \Lambda_{O3}.
\end{array}
\eqno(4.3)
$$
The VEVs  $\langle Y_e\rangle$,  $\langle Y_d\rangle$ 
and  $\langle \bar{E} \rangle$ contribute to 
the family gauge boson masses $m(A_i^j)$.
The VEVs $\langle Y_e\rangle$,  $\langle Y_d\rangle$ have
hierarchical structures, while  $\langle \bar{E} \rangle$ 
takes a structure proportional to a unit matrix.
Since it is not likely that the lightest family gauge boson mass $m(A_1^1)$ 
is smaller than $10^3$ GeV, we take
$$
\Lambda_{U3} \sim 10^3 \ {\rm GeV}. 
\eqno(4.4)
$$
Then, we may suppose
$$
10^3 \ {\rm GeV} \sim m(A_1^1) < m(A_2^2) < m(A_3^3) \sim 
10^4 \ {\rm GeV} ,
\eqno(4.5)
$$
because $m(A_3^3)$ is contributed from  $\langle Y_e\rangle$,  
$\langle Y_d\rangle$ and  $\langle \bar{E} \rangle$, 
while $m(A_1^1)$ is dominantly contributed only from
 $\langle \bar{E} \rangle \propto {\bf 1}$. 
Note that, usually, a scale of a family symmetry breaking 
cannot take a too low value, because such a low value contradicts 
phenomenology in  
the kaon physics. 
In contrast to the conventional models, in the present model, we can 
take a considerably low value of $\Lambda_{U3}$, because
the U(3) gauge bosons couple only to SU(2)$_L$ singlet down-quark
$d^c_i$, while they cannot couple to SU(2)$_L$ doublet quark
$(u_\alpha, d_\alpha)_L$.  
The value $m(A_1^1) \sim 10^3$ GeV is a value within our reach: 
The gauge boson $A_1^1$ can be observed via the characteristic 
decay $A_1^1 \rightarrow e^+ e^-$ (but no $\mu^+ \mu^-$)
\cite{KSY_11PLB} in $Z'$ search experiments at LHC and ILC.

On the other hand, for $\mu_R$,  
as a trial, let us assume
$$
\mu_R \sim M_{Pl} \sim 10^{19} \ {\rm GeV} ,
\eqno(4.6)
$$
where $M_{Pl}$ is the Planck mass. 
Then, from Eq.(3.21), we obtain
$$
\mu_R \langle {Y}_R^{\alpha\beta} \rangle 
\sim \langle \hat{Y}_u^{\alpha\gamma} \rangle
\langle \hat{Y}_e^{\gamma\beta} \rangle \sim 
\Lambda_{O3}^2 \sim \Lambda_{GUT}^2 ,
\eqno(4.7)
$$
which leads to the value of 
$\langle {Y}_R^{\alpha\beta} \rangle$ 
$$
\langle {Y}_R^{\alpha\beta} \rangle 
\sim (10^{16} {\rm GeV})^2/(10^{19} {\rm GeV})
\sim 10^{13} \, {\rm GeV} .
\eqno(4.8)
$$
Thus, we can obtain reasonable neutrino mass scale
which is consistent with Eq.(2.21).

Next, we discuss scales of the mass parameters $M_5$ and $M_{10}$.
The observed relations  
$m_\tau/ \langle H_d \rangle \sim m_b/\langle H_d \rangle
\sim 10^{-1}$ (we consider $\tan\beta \sim 10$) suggest
$$
M_5 \sim 10\, \Lambda_{U3} , 
\eqno(4.9)
$$
from Eq.(2.8), where we have regarded the VEVs of $Y_e$ and $Y_d$ 
as $\langle Y_e \rangle \sim \langle Y_d \rangle \sim \Lambda_{U3}$. 
On the other hand, the observed relation 
$m_t/\langle H_u \rangle \sim 1$ means 
$\langle \hat{Y}_u^{33}\rangle /{M}_{10} \sim 1$:
$$
 M_{10} \sim \langle \hat{Y}_u \rangle \sim \Lambda_{O3} 
\sim 10^{16} \ {\rm GeV} .
 \eqno(4.10)
$$

The assumption (4.9) is somewhat queer, because $M_5$ 
and $M_{10}$ are mass parameters of $({\bf 5}'_\alpha +
\bar{\bf 5}'_\alpha)$ and $(\overline{\bf 10}'_\alpha +
{\bf 10}'_\alpha)$, respectively, and both fields are triplets 
of O(3).
(Note that the constraints (4.9) and (4.10) are 
phenomenological ones, and they are not based on theoretical 
reasons.)  
In this paper, we regard $M_5$ and $M_{10}$ as merely 
parameters in the superpotential differently from 
the realistic masses of  $({\bf 5}'_\alpha +
\bar{\bf 5}'_\alpha)$ and $(\overline{\bf 10}'_\alpha +
{\bf 10}'_\alpha)$. 
Therefore, for the time being, the value of $\Lambda_{U3}$ is 
free, although we consider $\Lambda_{O3} \sim \Lambda_{GUT}
\sim 10^{16}$ GeV.

\begin{table}
\begin{center}
\begin{tabular}{c|ccccc} \hline
$M_{5}$ & $10^8$ GeV & $10^6$ GeV & $10^5$ GeV & $10^4$ GeV & $10^3$ GeV 
\\ \hline
$M_{10}= 10^{15}$ GeV  & $10.4$  & $8.2$ & $7.1$ & $6.0$ & $4.9$ \\
$M_{10}=10^{14}$ GeV & $7.1$ & $4.9$ & $3.8$ & $2.7$ & $1.6$ \\
\hline
\end{tabular}  
\end{center}
\begin{quotation}
\caption{
Value of $\alpha_5^{-1}$ at $\mu=\Lambda_{GUT}$ for typical values
of $M_5$ and $M_{10}$.
}
\end{quotation}
\end{table}

We investigate what value of $M_5$ is acceptable 
without blowing up the gauge coupling constants of 
the SU(3)$_c \times$SU(2)$_L\times$U(1)$_Y$ as seen 
in Table 3. 
Results are very sensitive to the value of $M_{10}$.
When we take $M_{10} = 10^{14}$ GeV and
$M_{5}= 10^8$ GeV, $10^6$ GeV and $10^3$ GeV,  
we obtain $\alpha_5^{-1}= 7.1$, $3.8$ and $1.6$, 
respectively ($\alpha_5$ is the SU(5) unification gauge
coupling constant) without blowing up. 
(We show an example of the behavior of the gauge coupling 
constants in Fig.3.)
Therefore, we can choose any low value of $\Lambda_{U3}$ 
(but $\Lambda_{U3} \ge 10^2$ GeV) as far as $M_{10} \ge 10^{14}$
GeV and $M_5 \ge 10^3$ GeV are concerned. 
However, a too low value of $\Lambda_{U3}$ is still not
unlikely.
In this paper, we suppose
$$
M_5 \sim 10^4\ {\rm GeV} , \ \ \ 
\Lambda_{U3} \sim 10^3\ {\rm GeV} .
\eqno(4.11)
$$

\begin{figure}[t!]
  \includegraphics[width=60mm,clip]{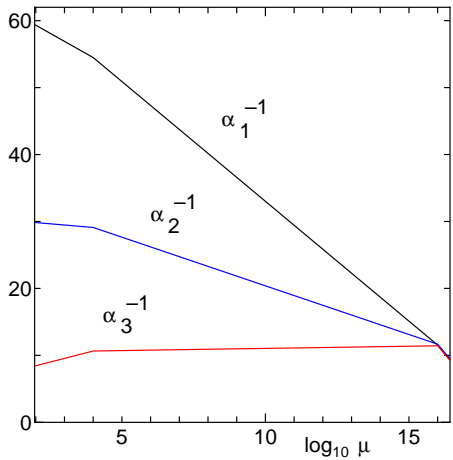}
  \caption{ Behavior of gauge coupling constants $\alpha_i^{-1}$
($i=1,2,3$) in the case of $M_{5}=10^{4}$ GeV 
and $M_{10}=10^{16}$ GeV.
For simplicity, we have neglected the SUSY breaking effects at 
$\mu \sim 10^3$ GeV in this figure. 
}
  \label{g-evl_SU5}
\end{figure}

As seen in Table 2, we have many U(3) non-singlet fields 
in the present model, so that the model does not give an 
asymptotic free theory. 
The evolution of the U(3) family gauge coupling constant 
$\alpha_F(\mu)$ is given by
$$
\frac{d}{d \log\mu} \alpha^{-1}_F(\mu) =\frac{1}{2\pi} 
\left( 9-\frac{1}{2} \sum \ell(R) \right),
\eqno(4.12)
$$
where $\ell(R)$ is an index of the representation $R$ of 
the group U(3).
The sum $\sum \ell(R)$ is given by $\sum \ell(R)=6$ 
for $\mu <\Lambda_{U3}$ and $\sum \ell(R)=15+15$ for
$\Lambda_{U3} < \mu < \Lambda_{GUT}$, where we do not
consider contribution from $\bar{\bf 5}^{\prime\prime}
+{\bf 5}^{\prime\prime}$ because they have masses of 
the order of $\Lambda_{GUT}$.
We find that $\alpha_F(\mu)$ does not blow up even in the
case of $\Lambda_{U3} =10^3$ GeV unless 
$\alpha_F(m_Z) > 0.033$. 
We show behavior of $\alpha_F(\mu)$ in a typical 
case with $\Lambda_{U3} = 2$ TeV and $\alpha_F(m_Z) =0.02$ in Fig.4.

\begin{figure}[t!]
  \includegraphics[width=60mm,clip]{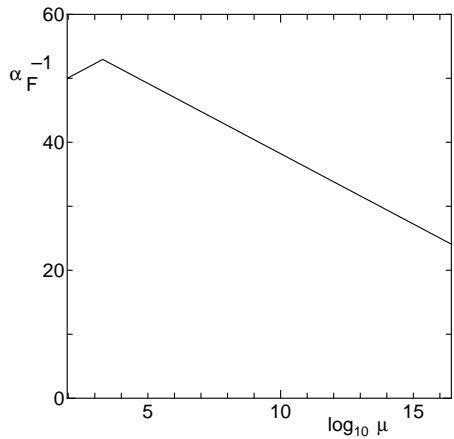}
  \caption{ Behavior of the inverse $\alpha_F^{-1}$ of the  
U(3) family gauge coupling constant 
in the case with $\Lambda_{U3}=2$ TeV  
and $\alpha_F(m_Z)=0.02$.
}
  \label{g-evl_U3}
\end{figure}

We do not discuss the behaviors of gauge coupling constants 
above $\mu=\Lambda_{GUT}$ because we have no scenario at 
$\mu > \Lambda_{GUT}$ at present.

\vspace{5mm}

{\large\bf 5. Concluding Remarks}

In conclusion, we have investigated a possibility that 
a family gauge symmetry U(3) has a comparatively low energy scale
by considering an SU(5) compatible yukawaon model with 
two family symmetries U(3)$\times$O(3).
Since all of yukawaons are SU(5) singlets, the existence of the yukawaons
do not affect the SU(5) GUT model, so that we can inherit the successful 
results in the SU(5) GUT.
However, the purpose of the present model is not to discuss problems
which are peculiar to the SU(5) GUT scenario. 
We optimistically consider that those problems will be resolved 
by considering further higher GUT groups (SO(10) or E$_6$, and so on) 
and/or an extra-dimension scenario.

In the present model, we have the following matter fields:
$$
(\bar{\bf 5}_i+{\bf 10}_\alpha+ {\bf 1}_\alpha) + 
(\bar{\bf 5}^{\prime\prime}_i+{\bf 5}^{\prime\prime\, i})
+(\bar{\bf 5}^{\prime}_\alpha+{\bf 5}^{\prime}_\alpha) 
+({\bf 10}^{\prime}_\alpha+\overline{\bf 10}^{\prime}_\alpha) ,
\eqno(5.1)
$$
where $i$ and $\alpha$ are indices of U(3) and O(3), respectively. 
The particles $(\bar{\bf 5}^{\prime\prime}_i+{\bf 5}^{\prime\prime\, i})$ 
and
$(\overline{\bf 10}^{\prime}+{\bf 10}^{\prime})_\alpha$ 
have masses of the orders of $\Lambda_{GUT} \sim 10^{16}$ GeV,
while $({\bf 5}^{\prime}+\bar{\bf 5}^{\prime})_\alpha$ have
masses of the order of $10^{4}$ GeV.
The U(3) family symmetry is broken at $\mu= \Lambda_{U3} 
\sim 10^{3}$ GeV. 

The most notable result is that we have been able to consider 
a double seesaw mechanism for up-quark mass generation as shown 
in Fig.2 by introducing O(3) family symmetry. 
(If we consider U(3)$\times$U(3) family symmetries, we cannot obtain the
effective Yukawa interaction (2.12).)
As a result, the $Y_d$-$\hat{Y}_u$ corresponding has 
been improved as seen in Eq.(1.15). 
Also, by considering that the fundamental yukawaon 
$\Phi_e$ is transformed a triplet (doublet + singlet) of
a permutation symmetry as defined in Eq.(3.3), our model
without a cutoff $\Lambda$ can take more simple forms. 

In this paper, we did not give numerical results on the 
basis of the present model, because the phenomenology is 
almost the same as the previous model \cite{O3_09PLB}. 
Phenomenology for the family gauge bosons with the 
scale $10^3$ GeV will be given elsewhere.

\vspace{10mm}
{\Large\bf Acknowledgments}   

The author would like to thank T.~Yamashita for helpful conversations. 
The work is supported by JSPS 
(No.\ 21540266).

%


\end{document}